\documentclass[11pt]{article}

\usepackage{amssymb,amsmath,epsfig,fullpage}

{\bfseries}{\itshape}

\begin{document}

\title{Compact pairwise models for epidemics with multiple infectious stages on degree heterogeneous and clustered networks}

\author{N. Sherborne}

\author{N. Sherborne, \hspace{0.5cm}K.B. Blyuss\thanks{Corresponding author. Email: k.blyuss@sussex.ac.uk}, \hspace{0.5cm}I.Z. Kiss 
\\\\ Department of Mathematics, University of Sussex, Falmer,\\
Brighton, BN1 9QH, United Kingdom}

\maketitle
	
\begin{abstract}
This paper presents a compact pairwise model that describes the spread of multi-stage epidemics on networks. The multi-stage model corresponds to a gamma-distributed infectious period which interpolates between the classical Markovian models with exponentially distributed infectious period and epidemics with a constant infectious period. We show how the compact approach leads to a system of equations whose size is independent of the range of node degrees, thus significantly reducing the complexity of the model. Network clustering is incorporated into the model to provide a more accurate representation of realistic contact networks, and the accuracy of proposed closures is analysed for different levels of clustering and number of infection stages. Our results support recent findings that standard closure techniques are likely to perform better when the infectious period is constant.
\end{abstract}

\section{Introduction}

Mathematical models of infectious diseases have proven to be an invaluable tool in understanding how diseases invade and spread within a population, and how best to control them \cite{anderson1991infectious,diekmann2012mathematical,RevModPhys.87.925}. 

Given a good understanding of the biology of the disease and of the behaviour and
interaction of hosts, it is possible to develop accurate models with good predictive power,
which provide the means to develop, test and deploy control measures to mitigate the negative
impacts of infectious diseases, a good example being influenza \cite{ferguson2006strategies}. However, as has been
highlighted by the recent Ebola outbreak in West Africa \cite{chowell2014transmission},
models can be very situation-specific and can become highly sophisticated or complex
depending on intricacies of the structure of the population and the characteristics of the
disease.

In the last few decades the use of networks to describe interactions between individuals has been an important step change in modelling and studying disease transmission \cite{keeling1999,danon,keeling2005networks,RevModPhys.87.925}. There is now overwhelming empirical evidence
that in many practical instances individuals interact in a structured and selective way, e.g. in the case of sexually transmitted diseases \cite{liljeros2001web}. Thus, the well-mixed assumption of early compartmental models \cite{SIR} has to be relaxed or models need to be refined by including multiple classes and mixing between classes. However, in some cases a network representation could be more realistic than a description based on compartmental models. Conventionally, nodes in network-based models represent individuals, and the edges describe connections between people who have sufficient contact to be able to transmit the disease \cite{keeling2005networks,danon,RevModPhys.87.925}. The total number of edges a node has is known as its degree, and the frequency of nodes with different degrees is determined by a specific \emph{degree distribution} $P(k)$ which can either be empirically measured or given theoretically. In either case $P(k)$ is the probability of a randomly chosen node having degree $k$. Early network models often assumed regular networks where all nodes have the same degree, or well-studied networks from graph theory, such as the Erd{\H o}s-R{\'e}nyi random graphs \cite{erdds1959random}. However, empirical research showed that real biological, social or technological networks do not conform to such idealised models. In fact, many studies on human interactions ranging from sexual contact networks \cite{liljeros2001web} to using the travel of banknotes as an indicator of human activity \cite{brockmann2006scaling}, or even internet connectivity \cite{webconnect} have observed \emph{wide-tail distributions}, with the majority of nodes having a low number of contacts, and a few nodes in the network having a much higher degree. This structure is most closely approximated by scale-free networks described by a power-law degree distribution $P(k) \sim k^{-\alpha}$ with some positive exponent $\alpha$, which for most accurately described human contact patterns lies in the range $\alpha \in [2,3]$ (see, for example, \cite{pastor2001epidemic}). The impact of contact heterogeneity on the spread of epidemics is significant, and studies have highlighted the disproportionate role which may be played by a few highly-connected nodes \cite{james2007event}.

Another striking feature of real social contact patterns is the presence of small and highly-interconnected groups which occur much more frequently than if edges were to be distributed at random. This is known as {\it clustering}, and its presence in empirical data \cite{newman2001random,foster2011clustering} has driven the need to consider network models that include this feature. Perhaps, one of the most well-known and parsimonious theoretical models with tuneable clustering is the small-world network \cite{watts1998collective}, where nodes are placed on a ring, and the network is dominated by local links to nearest neighbours with a few links rewired at random, which means that the average path length is not too large and comparable to that found in equivalent random networks. For a summary of numerous alternative algorithms that can be used to generate clustered networks see, for example, \cite{green2010large} or \cite{ritchie2014higher}. It is well known that modelling epidemic spread on such networks is more challenging, although some models have successfully incorporated clusterings \cite{miller2009percolation,karrer2010random,volz2011effects,ritchie2015beyond} (and references therein). However, it is often the case that such models only work for networks where clustering is introduced in a very specific way, e.g. by considering non-overlapping triangles or other subgraphs of more than three nodes. 

Besides the details of the network structure, another major assumption that significantly reduces the mathematical complexity of models and makes them amenable to analysis with mean-field models of ordinary differential equations and tools from Markov chain theory is the assumption that the spreading/transmission of infection and recovery processes are Markovian. However, it has long been recognised that this is often not the case, and, for example, the infectious periods are typically far from exponential, and, perhaps, are better described by a normal-like or peaked distribution \cite{gough1977estimation,lloyd2001realistic,wearing2005appropriate}. Modelling non-Markovian processes can be challenging and often leads to delay differential or integro-differential equations that are much more difficult to analyse. Recently, \cite{kiss2015pairwise} have put forward a generalisation of a pairwise model for Markovian transmission with a constant infectious period for a susceptible-infected-recovered (SIR) dynamics. The resulting model is a system of delay differential equations with discrete and distributed delays which makes it possible to gain insight into how the non-Markovian nature of the recovery process affects the epidemic threshold and the final epidemic size. Other important recent research in this direction includes the message passing formalism \cite{karrer2010message, wilkinson2014message} and an approach based on renewal theory \cite{cator2013susceptible}.

In light of the importance of the above-mentioned network properties (i.e. degree heterogeneity and clustering) and the non-Markovian nature of the spreading and/or recovery processes driving the epidemics, in this paper we generalise our recent research on a multi-stage $SIR$ epidemic model \cite{sherbs} and focus on modelling a Markovian spreading process with gamma-distributed infectious period on networks that account for heterogeneous degree distribution and clustering. This is achieved within the framework of pairwise models \cite{keeling1999}, and we show that the additional model complexity induced by degree heterogeneity and non-Markovian recovery can be effectively controlled via a reduction procedure proposed by \cite{simon2015super}. This allows one to derive an approximate deterministic model that helps numerically determine the time evolution of the epidemic and the final epidemic size. Moreover, the model allows us to gain insights into the interactions of the three main model ingredients, namely, degree heterogeneity, clustering and non-exponential recovery and the agreement between the model and the stochastic network simulation. The paper is organised as follows. In the next section we derive a compact pairwise model for unclustered networks whose size is independent of the range of degrees and derive and discuss some analytical results for this model. All results are validated by comparing the numerical solution of the pairwise model to results from direct stochastic network simulation. In Section \ref{sec:clustered}, we investigate the case when the same epidemic unfolds on clustered networks. The corresponding pairwise model is derived, and we discuss the extra complexities necessary to more accurately approximate the spread of the disease. More importantly, we investigate how clustering and the non-Markovian recovery affect the agreement between the pairwise model and simulations. Finally, in Section \ref{sec:postmortem} we conclude with a discussion of our results and future work.

\section{Disease dynamics in the absence of clustering}
\label{sec:basic}

As a first step in the analysis of the spread of epidemics on unclustered networks, we introduce the necessary concepts from multi-stage infections and pairwise models \cite{sherbs}. In the $SI^KR$ model, once a susceptible individual $S$ becomes infected, they progress through $K$ equally infectious stages denoted as $I^{(i)}$, $1\leq i\leq K$. The transition rates between successive stages are given by $K\gamma$. Thus, in simulation the times spent in each of the $K$ stages are independent exponentially distributed random numbers. The total time of infection is, therefore, the sum of $K$ exponential distributions, which is a gamma distribution with the mean time of $\gamma^{-1}$ \cite{durr}. In order to describe the dynamics of an epidemic we consider the state of the nodes in the network and the edges connecting them. Since a susceptible individual can only become infected upon a transmission across an $S-I^{(i)}$ link we need to consider the expected number of edges connecting susceptible and infected individuals (in any of the $K$ stages) at time $t$ over the whole network, to be denoted as $[SI^{(i)}](t)$. Here we have taken $[SI^{(i)}]$ independently of the degrees of the nodes in state $S$ and $I^{(i)}$, i.e. $[SI^{(i)}] = \sum_{a,b} [S_aI_b^{(i)}]$ where $a$ and $b$ denote the degrees in the range between the minimum and maximum degrees in the network, denoted as $k_{min}$ and $k_{max}$, respectively. This definition applies to all pairs, i.e. $[AB]$ stands for the population level count of all $A-B$ edges taken across all possible connections between nodes of different degrees; 
\[
[AB] = \sum_{a,b} [A_aB_b] ,\quad \mbox{and} \quad A,B \in \{S, I_1, I_2, \ldots, I_K, R\}:=\mathbb{S}.
\] 
Here and henceforth $\mathbb{S}$ will denote the set of all possible states for a node. The expected number of $S-S$ edges depends on the expected number of $S-S-I^{i}$ triples, with this being the case for other edge types as well. To break the dependency on higher order moments, closure relations must be introduced which allow us to approximate the number of triples using the number of pairs and nodes in different states \cite{keeling1999}.

\begin{figure}
	\hspace{-1cm}
	\epsfig{file=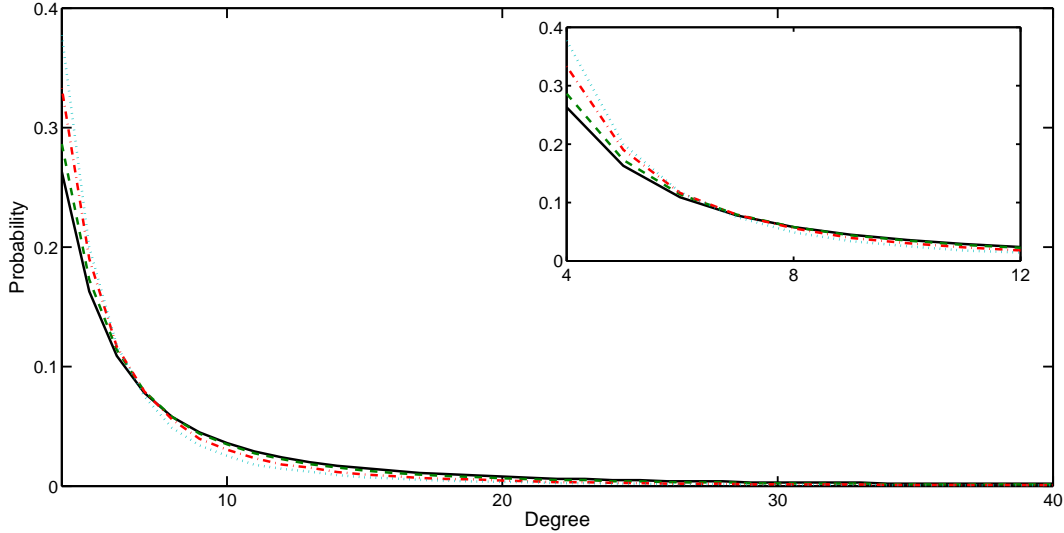, width=17cm}
	\caption{(Colour online) The results of testing the relation between the degree distribution $P(k)$ and the distribution of susceptible nodes $s_k$ over time for a truncated scale-free network. The black line represents the degree distribution $P(k)$ that coincides with $s_k$ at $t=0$, and the green, red and light blue lines represent $s_k$ at times $10$, $15$ and $20$, respectively from $100$ simulations of the epidemic. Note that all lines show the same qualitative behaviour.}
	\label{fig:lin_evidence}
\end{figure}

We begin our analysis by considering the simpler case where the contact network has a locally tree-like structure characterised by zero clustering. The Markovian, or single stage, pairwise model has been proven to be exact prior to closure~\cite{taylor2012markovian}, and the approach can be extended to a $SI^KR$ multi-stage model. In order to obtain a pairwise model for the $SI^{K}R$ dynamics for unclustered and degree heterogeneous networks, we start with the unclosed model for a general $K$-stage disease and describe an \emph{a priori} method to derive a new set of closures at the level of triples. It should be noted that our approach resembles that used in recent works of \cite{simon2015super} and \cite{house2011insights}. The system describing the dynamics of a $K$-stage disease has the following form \cite{sherbs}
\begin{align}
&\dot{[S]}   = -\tau [SI], \nonumber \\
&\dot{[I^{(1)}]} = \tau [SI] - K\gamma [I^{(1)}], \nonumber\\
&\dot{[I^{(j)}]} = K\gamma [I^{(j-1)}] - K\gamma [I^{(j)}], \quad \text{for} \quad j=2, 3, \ldots, K, \nonumber\\
&\dot{[SS]}= -2\tau[SSI], \label{eq:ex_pair}\\
&\dot{[SI^{(1)}]}= -(\tau+K\gamma)[SI^{(1)}] + \tau\left([SSI] - [ISI^{(1)}]\right), \nonumber\\
&\dot{[SI^{(j)}]}= -(\tau+K\gamma)[SI^{(j)}] + K\gamma[SI^{(j-1)}] -\tau [ISI^{(j)}], \quad \text{for} \quad j=2, 3, \ldots, K,\nonumber\\
&\dot{[SR]} = -\tau [ISR] + K\gamma [SI^{(K)}], \nonumber
\end{align}
where $\tau$ is the per-link disease transmission rate, and the terms without superscripts represent summation over all infected compartments, i.e. $[SI]=\sum_{i=1}^{K}[SI^{(i)}]$, $[SSI]=\sum_{i=1}^{K}[SSI^{(i)}]$ and $[ISI^{(j)}]=\sum_{i=1}^{K}[I^{(i)}SI^{(j)}]$. While the above equations do not seem to account separately for the degrees of the nodes, we will show that it is possible to keep such a system and include all the information about the degree distribution in a new closure relation at the level of pairs. The closure for this model can be obtained by first considering the classical triple closure for a regular network proposed by \cite{keeling1999}
\begin{equation}
	[XSI^{(i)}] \approx \frac{n-1}{n} \frac{[XS][SI^{(i)}]}{[S]},
	\label{eq:classic_closure}
\end{equation}
where $n$ is the degree of every node in the network (and thus also the mean degree), and $X\in \mathbb{S}$. The derivation of a new closure for heterogeneous networks starts from noting that closure (\ref{eq:classic_closure}) depends on the degree of the middle node, which allows us to write
\begin{equation}\label{eq:mid_closure}
	\displaystyle{[XS_jY] \approx \frac{j-1}{j} \frac{[XS_j][S_jY]}{[S_j]},} \quad X, Y \in \mathbb{S},
\end{equation}
for a susceptible node of degree $j$, with $j\in [k_{min}, k_{max}]$. To make further progress, one can use the approximation used by \cite{eames2002modeling},
\begin{equation}\label{eq:secondapprox}
	\displaystyle{[S_jY] \approx [SY] \frac{j[S_j]}{\sum_{m=k_{min}}^{k_{max}} m[S_m]}.}
\end{equation}
This assumes that the number of $S_j-Y$ pairs is approximately equal to the number of $S-Y$ pairs (regardless of node degree) multiplied by the fraction of $S$ nodes with degree $j$. Substituting this approximation into (\ref{eq:mid_closure}) yields
\begin{equation}
	[XS_jY] \approx [XS][SY] \frac{j(j-1)[S_j]}{T_1^2},
	\label{eq:XSjY}
\end{equation}
where
\[
T_1:=\displaystyle{\sum_{m=k_{min}}^{k_{max}} m[S_m] = [SS] + \sum_{i=1}^K[SI^{(i)}] + [SR]}
\]
denotes the total number of edges emanating from susceptible nodes. The second expression for $T_1$ above follows directly from the pairwise model~(\ref{eq:ex_pair}) and explains the need for explicitly including an equation for $[SR]$. Taking the sum of all triples in (\ref{eq:XSjY}) over all degrees $j$ gives
\begin{equation}
	[XSY] = \sum_{j=k_{min}}^{k_{max}}[XS_jY] \approx [XS][SY] \frac{T_2-T_1}{T_1^2},
	\label{eq:sum_close}
\end{equation}
with
\[
T_2=\displaystyle{\sum_{m=k_{min}}^{k_{max}} m^2[S_m].}
\]
Unfortunately, $T_2$ cannot be expressed in a closed form from the solution of system (\ref{eq:ex_pair}). However, it should be possible to estimate the degree distribution of susceptible nodes \cite{shkarayev2014epidemics, simon2015super}. This distribution is given by 
\[
s_k := [S_k]/[S],
\]
and has the mean
\[
n_S = T_1/[S].
\]
\cite{simon2015super} have shown by means of numerical simulations that the (dynamic) degree distribution of susceptible nodes is proportional to the degree distribution $P(k)$. Numerical simulations shown in Fig.~\ref{fig:lin_evidence} demonstrate that despite being entirely heuristic, this relation between the two distributions holds for all the different networks it was tested on. We use this linear relationship between $s_k$ and $P(k)$ in order to derive a compact model. A brief explanation is given below, for the full method one can refer to \cite{simon2015super}. As they will be needed later, we first introduce the moments of the degree distribution $P(k)$, namely,
\begin{equation*}
	\displaystyle{n_i= \frac{\sum_{m=k_{min}}^{k_{max}} m^i N_m}{N}=\sum_{m=k_{min}}^{k_{max}} m^i P(m).}
\end{equation*}
It is easy to see that 
\[ T_2 = [S] \sum_{m=k_{min}}^{k_{max}} m^2 s_m, \]
and so our goal is to find an estimate for $s_k$. Introducing a new variable $q_k = s_k/P(k)$
linearity enforces the following relation for all $k \in [k_{min}, k_{max}]$
\[ \frac{q_k - q_{k_{min}}}{k - k_{min}} = \frac{q_{k_{max}} - q_{k_{min}}}{k_{max} - k_{min}}. \]
By manipulating this equation one can identify a relation between $s_k$ and $P(k)$; namely
\begin{equation}
	s_k = \dfrac{\left(k - k_{min}\right)q_{k_{max}} + \left(k_{max} - k\right)q_{k_{min}}}{k_{max} - k_{min}} P(k).
	\label{eq:31}
\end{equation}
Since the sum of all $s_k$'s is one, and the distribution has the mean $n_S$, it is then possible to recast $q_{k_{min}}$ and $q_{k_{max}}$ in terms of the known quantities $n_1$, $n_2$, $n_3$ and $n_S$. Feeding these back into (\ref{eq:31}) gives an estimate for $s_k$, and thus $T_2$. Using this estimate we arrive at the following relation
\begin{equation*}
	\frac{T_2-T_1}{T_1^2} \approx \frac{1}{n_S^2[S]}\left(\frac{n_2(n_2 - n_1n_S) + n_3(n_S - n_1)}{n_2 - n_1^2} - n_S\right).
\end{equation*}
This gives the closure for the heterogeneous compact pairwise $SI^KR$ model~(\ref{eq:ex_pair}) in the form
\begin{align}\label{eq:new_close}
	[XSY]\approx \zeta(t)\frac{[XS][SY]}{[S]},
\end{align}
where 
\begin{equation}
	\zeta(t)=\frac{n_2(n_2 - n_1n_S) + n_3(n_S - n_1)}{n_S^2\left(n_2 - n_1^2\right)} - \frac{1}{n_S}.
	\label{eq:zeta}
\end{equation}
It is evident that the range of degrees and the degree distribution have been implicitly accounted for in the closure relation, thus allowing us to work with a set of equations whose size is independent of the range of degrees. In other words, regardless of the exact nature of the contact network we will only ever need $2K+3$ equations in~(\ref{eq:ex_pair}) to model the epidemic. This is due to all of the information about the degree distribution being included in $\zeta(t)$. In the special case of regular contact networks, where every node has the same degree $n$, one has that $n_S = n_1=n$, $n_2 = n^2$ and $n_3 = n^3$, hence $\zeta(t)$ reduces to
\[
\zeta=\frac{n-1}{n},
\]
and the closure reverts back to the simpler version given in~(\ref{eq:classic_closure}).

\subsection{Numerical simulation results}
\label{sec:basic_sims}

\begin{figure} 
	\hspace{-1cm}
	\epsfig{file=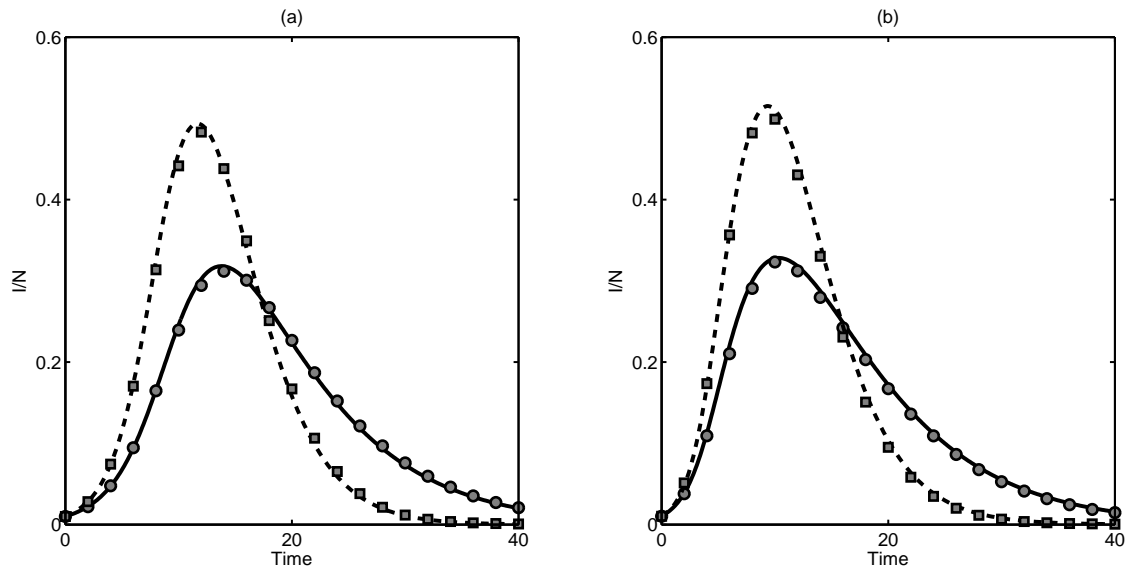,width=18cm}
	\caption{Dynamics of epidemics spreading on unclustered networks of 1000 nodes with (a) bimodal degree distribution with an even split of nodes having degrees $4$ or $12$, and (b) truncated scale-free degree distribution $P(k) \sim k^{-\alpha}$ bounded by $k_{min}=4$, $k_{max}=60$, and with $\alpha =2.5$, and the mean degree of around $8$. For both topologies, the simulations are performed for $K=1$ (black line, circles) and $K=4$ (dashed line, squares). Lines show the solution of the pairwise model (\ref{eq:ex_pair}) with the closures given in (\ref{eq:new_close}), and symbols correspond to stochastic network simulation. Other parameter values are $\tau=0.07$, $\gamma=0.15$.}
	\label{fig:unclustered}
\end{figure}

In order to test the effectiveness of model~\ref{eq:ex_pair}) with closure~(\ref{eq:new_close}), we compare its output to numerical simulation of epidemics spreading on networks with bimodal and truncated scale-free degree distributions, with both types of networks being constructed using the configuration model \cite{bender1978asymptotic}. For bimodal networks, all nodes have degree $k_1$ or $k_2$, and the proportion of nodes with degrees $k_1$ and $k_2$ in the network; each node is then given either $k_1$ or $k_2$ half-edges which are connected to other half-edges at random to create the edges. The generation of truncated scale-free networks begins by choosing bounds of minimum and maximum degree $k_{min}$ and $k_{max}$. One then generates a power law distribution with a chosen exponent $\alpha$ and samples the normalized probability of a node having degree $k \in [k_{min}, k_{max}]$, after which half-edges are drawn and connected at random. If the total number of half-edges is odd, one is removed at random, the effect of which is small and diminishes rapidly as the total number of nodes $N$ grows. Each simulation begins with a single infected individual, and the time is rescaled to zero after the number of infected individuals reaches ten, when counted across all compartments. The results of these tests are presented in Fig.~\ref{fig:unclustered}, which show the comparison of an average of 100 simulations (consisting of 20 simulations for five different random networks with the same topology) and the output from the pairwise model~(\ref{eq:ex_pair}).
Figure \ref{fig:unclustered} shows that increasing the number of infectious stages leads to a more rapid spread of the disease with higher peak prevalence, despite the mean duration of infection remaining unchanged. This suggests that the lead time to implement any control measures is much shorter than estimates based on standard models where recovery is Poisson would suggest. This behaviour was also observed in the case of homogeneous populations \cite{sherbs}. We further note that for the same parameters of the disease dynamics, the trend of faster growth is even more profound for scale-free networks. This effect can be attributed to the influence of a small number of highly connected nodes; these individuals are at greater risk of receiving infection, and also have a much greater capacity to spread the disease, thereby causing a rapid increase in the number of new infections. This also has a significant impact on the threshold parameter which describes the point at which an epidemic occurs, as will be discussed later.

\subsection{Characteristics of the multi-stage compact model}
\label{sec:analysis}

Now that the system of pairwise equations (\ref{eq:ex_pair}) with closures given in (\ref{eq:new_close}) has been shown to accurately match simulations for a range of networks, we focus on deriving  analytical results from this model. The first quantity we consider is the \emph{transmissibility} of the disease, defined as the probability of the disease being successfully transmitted across a given $S-I$ link, when considered in isolation. To compute this quantity, we recall that the recovery times are now gamma-distributed. For a successful infection attempt to occur across an $S-I$ link, the infection must be transmitted before the infected node recovers, hence, it can be computed as follows \cite{sherbs},
\begin{equation}
	\widetilde{\tau} :=\int_0^{\infty} \left(1 - e^{-\tau x}\right)\frac{1}{(K-1)!}(K\gamma)^K x^{K-1} e^{-(K\gamma)x}dx
	= 1 - \left(\frac{K\gamma}{\tau + K\gamma}\right)^K.\label{eq:tilde_tau}
\end{equation}

Although this estimate for the probability of transmission provides some indication of how likely a major epidemic is, it does not, however, take into account the heterogeneity in the network structure. To identify a threshold parameter that can indicate whether an epidemic will occur, we perform a linear stability analysis of the disease-free equilibrium (DFE)  with $[S]= N$, $[SS]=n_1N$, $[I^{(j)}]=[SI^{(j)}]=[SR]=0$, $j=1,2 \ldots,K$ of system (\ref{eq:ex_pair}) with the closure given in (\ref{eq:new_close}). If the DFE is stable, then any small outbreak will die out. The stability of the DFE is determined by eigenvalues of the Jacobian matrix $\mathcal{J} \in \mathbb{R}^{(2K+2)\times (2K+2)}$ ($[SR]$ can be safely excluded as it only introduces a further row and column of zeros). Due to the nature of the system, $\mathcal{J}$ can be recast in the block form $\mathcal{J} = \left(\begin{array}{cc}
A & B \\
C & D \end{array}\right),$ where $A$ is a lower-diagonal $(K+2)\times(K+2)$ matrix, $B$ is a $(K+2)\times K$ matrix, $C$ is a zero $K \times (K+2)$ matrix, and $D$ is a $K\times K$ matrix. This simplifies the calculations, since the characteristic equation can be rewritten as the product of diagonal elements of the matrix $A$ multiplied by the determinant of the matrix $D$, i.e.
\begin{equation*}
\lambda^2(\lambda+K\gamma)^K
\left|\begin{array}{ccccc}
\tau n_1 \zeta(0) - K\gamma - \tau - \lambda& \tau\zeta(0) 								& & \ldots & \tau\zeta(0) \\
K\gamma							& -K\gamma - \tau - \lambda & 0 & \ldots & 0			 \\
0										&	K\gamma										& \ddots & \ddots & \vdots \\
\vdots							& 0													& \ddots & \ddots & 0			\\
0										& \ldots										& 0			 & K\gamma & -K\gamma - \tau - \lambda 
\end{array}\right| =0.
\end{equation*}
This equation is similar to the one analysed by \cite{sherbs}. At time $t=0$ note that $n_S = n_1$, thus from (\ref{eq:zeta}) $\zeta(0) = (n_2-n_1)/(n_1^2)$. By considering the conditions under which the maximum eigenvalue changes its sign, it is possible to identify a threshold parameter
\begin{equation}\label{eq:threshold}
	\mathcal{R}:= n_1 \frac{n_2-n_1}{n_1^2} \widetilde{\tau}= \frac{n_2-n_1}{n_1} \widetilde{\tau},
\end{equation}
such that for $\mathcal{R} < 1$ the epidemic will die out, and for $\mathcal{R} > 1$ the epidemic will develop in the deterministic model~(\ref{eq:ex_pair}). This threshold translates to stochastic simulations, however, there is still a small possibility that an early disease die-out can occur even when $\mathcal{R} > 1$. Similarly, small epidemics may occur in some cases where $\mathcal{R}<1$. It is important to note that although $\widetilde{\tau}$ emerges directly from the linear stability analysis, identifying it as the transmissibility restores the conventional interpretation of the threshold for epidemic spread as the expected number of secondary infections caused by a single infected individual in a fully susceptible population. In this way, our findings agree with the literature (see, for example, \cite{diekmann1998deterministic}).

An interesting result can be reached by considering $\mathcal{R}$ in the case of a scale-free distribution with $P(k) \sim k^{-\alpha}$ where $\alpha \le 3$. In this case, unless $P(k)$ is truncated, higher moments $n_2$, $n_3$ of the degree distribution are not defined as the population size tends to infinity, and, hence, as the population size grows, the threshold parameter $\mathcal{R}$ will diverge for any non-trivial choice of the disease parameters $\tau$, $\gamma$ and $K$. Under these circumstances, the network topology dominates the dynamics of disease, and unless the contact structure can be altered or influenced, the disease will always spread through the population. This conclusion has been reached before in other models \cite{pastor2001epidemic}.

\begin{figure} 
	\hspace{-1cm}
	\epsfig{file=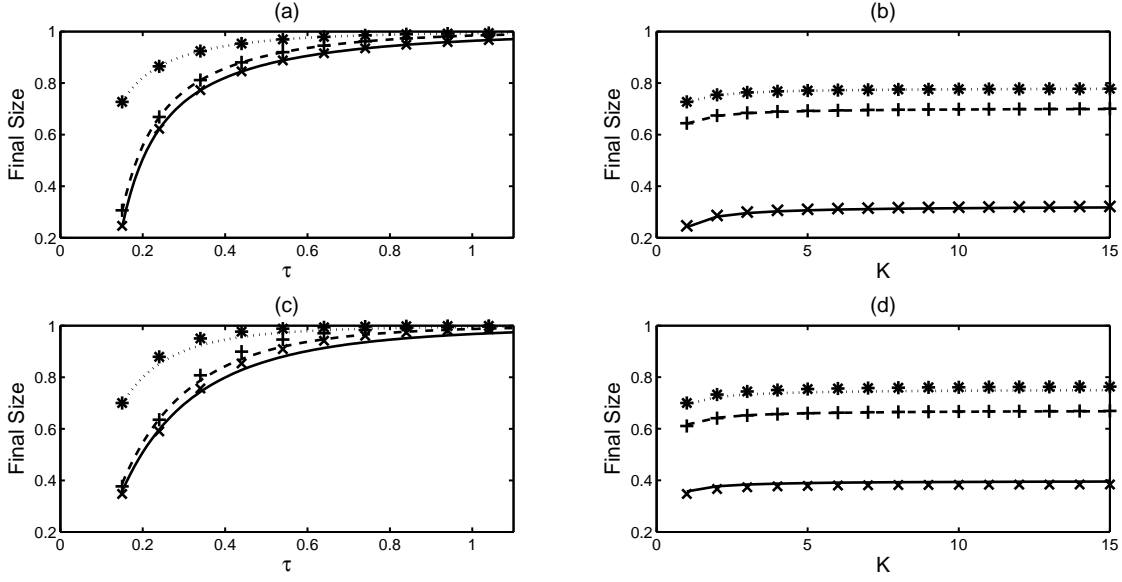,width=18cm}
	\caption{Comparison of the final epidemic size as determined by equations (\ref{eq:final_size1}) and (\ref{eq:final_size_consistency}) (lines) and solutions of the pairwise model (\ref{eq:ex_pair}) (markers) for bimodal (a)-(b) and truncated scale-free networks (c)-(d), respectively. The parameter values are: (a), (c) $K=1$, $\gamma=1$ (solid line, crosses), $K=4$, $\gamma=1$ (dashed line, pluses), $K=1$, $\gamma=0.5$ (dotted line, stars); (b), (d) $\tau=0.15$, $\gamma=1$ (solid line, crosses), $\tau=0.25$, $\gamma=1$ (dashed line, pluses), $\tau=0.15$, $\gamma=0.5$ (dotted line, stars).}
	\label{fig:newman_comparison}
\end{figure}

Since we are studying the spread of epidemics in a closed population, every epidemic will reach an end when there are no more infected individuals, at which point every member of the population is either still susceptible or in the removed class. To quantify the severity of an epidemic, it is instructive to look at the proportion of the population who will become infected over the entire lifetime of the epidemic; this quantity is known as the \emph{final epidemic size}. In principle, it may be possible to manipulate the equations in (\ref{eq:ex_pair}) with the newly derived closure approximation (\ref{eq:new_close}) to find first-integral-like relations and thus find an expression for the final epidemic size \cite{keeling1999, sherbs}. However, by considering the final epidemic size problem using a bond percolation model,~\cite{newman2002spread} showed that it is possible to obtain an exact result for the mean final epidemic size. Based on the generating function for the degree distribution $G_0(x):=\sum_{k}p_k x^k$, where $P(k)=p_k$, the generating function for the excess degree distribution $G_1(x) = \frac{1}{n_1}G_0'(x)=\frac{1}{n_1} \sum_{k}k p_k x^{k-1}$, and the transmissibility, which for our model is given by $\widetilde{\tau}$ in~(\ref{eq:tilde_tau}), the final epidemic size is given by \cite{newman2002spread}:
\begin{equation}
	\mathcal{R}_{\infty} = 1 - G_0(1 +(\theta-1)\widetilde{\tau}) = 1 - \sum_{k} p_k (1 +(\theta-1)\widetilde{\tau})^k, \label{eq:final_size1}
\end{equation}
where $\theta$ is the unique solution in $(0,1)$ of the following equation
\begin{equation}
	\theta = G_1(1 + (\theta-1)\widetilde{\tau}) = \frac{1}{n_1}\sum_{k} p_k (1 +(\theta-1)\widetilde{\tau})^{k-1}. \label{eq:final_size_consistency}
\end{equation}
Newman's work has been revisited by~\cite{kenah2007second}, and whilst they showed that the distribution of final sizes suggested by Newman's original work was incorrect for non-constant infectious periods the mean final epidemic size given by~(\ref{eq:final_size1}) and~(\ref{eq:final_size_consistency}) is correct.

Figure \ref{fig:newman_comparison} shows the comparison of the final epidemic size results based on equations (\ref{eq:final_size1}) and (\ref{eq:final_size_consistency}) to results from the numerical solution of the new pairwise model (\ref{eq:ex_pair}), and the agreement is excellent. It is noteworthy that in all cases the final epidemic size behaves as expected with respect to the disease parameters, i.e. a higher (lower) transmission rate $\tau$ results in a larger (smaller) final epidemic size, the mean duration of infection ($\gamma^{-1}$) has a similar effect, and a tighter distribution of the infectious periods (higher $K$) increases the predicted final epidemic size. Furthermore, a careful comparison of bimodal and truncated scale-free networks shows that having a broader degree distribution leads to certain differences in the dynamics. Namely, for relatively low transmission rates, epidemics of measurable size are predicted in truncated scale-free networks but not necessarily for the bimodal distribution. However, as the transmissibility grows (either through increasing $\tau$ or $K$) there comes a point where the final epidemic size becomes larger for the bimodal network. This is likely due to the large number of low-degree nodes in truncated scale-free networks, it is difficult for any epidemic to reach these nodes, even once highly-connected nodes have been infected.

\subsection{Limiting cases}
\label{sec:Kinf}

\begin{figure}
	\hspace{-1cm}
	\epsfig{file=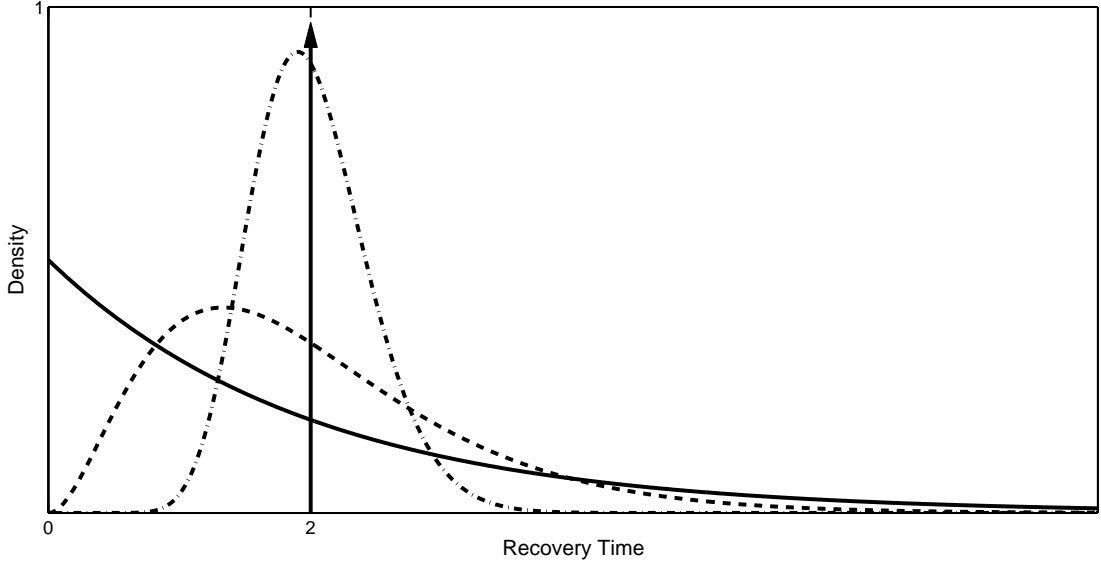,width=18cm}
	\caption{The distribution of infectious periods in the Markovian case of $K=1$ (solid), for $K=3$ (dashed), $K=20$ (dash-dotted), and, the Dirac delta distribution corresponding to $K=\infty$. The mean infectious period is equal to $2$ for all four distributions.}
	\label{fig:limit_cases}
\end{figure}

It is instructive to look at the behaviour of model~(\ref{eq:ex_pair}) in two particular limits of the number of infectious stages. When $K=1$, model (\ref{eq:ex_pair}) reverts to the classical Markovian pairwise model which has been thoroughly studied \cite{eames2002modeling, house2011epidemic}. As the number of stages increases, the shape of the distribution for the infectious period changes, as shown in Fig.~\ref{fig:limit_cases}. For larger $K$ one can see that the distribution grows tighter around the mean, which is kept constant at $\gamma^{-1}$ due to the particular formulation of the model, and there is also much less variation in the duration of infection. The limiting case of $K \to \infty$ results in the infected period having a Dirac delta distribution $\delta\left( t-\gamma^{-1}\right)$ around the mean infectious period. It has been recently shown that this case can be accurately described by a system of pairwise delay differential equations (DDEs) for homogeneous populations \cite{kiss2015pairwise}, in which case the above-mentioned concept of transmissibility is also applicable. In this case, the transmission process is still Markovian (thus the spreading process is characterised by the probability density function $\tau e^{-\tau t}$), however, the infectious period is now constant, hence the probability of the infected node recovering is given by $\xi(t)$, where
\[ \xi(t) = \begin{cases}
			0 \quad \text{if} \quad 0 \le t < \gamma^{-1}, \\
			1 \quad \text{if} \quad t \ge \gamma^{-1}.
\end{cases} \]
Under these circumstances the transmissibility for a disease with a constant infectious period is given by
\[
\displaystyle{\widetilde{\tau}_{\rm const.} = \int_0^{\infty}\tau e^{-\tau x} \xi(x) dx= 1 - e^{-\tau/\gamma}.}
\]
It is easy to show that taking the limit $K \to \infty$ in (\ref{eq:tilde_tau}) yields the same result, i.e.
\[
\lim\limits_{K\to\infty} \widetilde{\tau}= 1 - e^{-\tau/\gamma}.
\] 
This suggests that results for the final epidemic size and the threshold parameter $\mathcal{R}$ for the case of a constant infectious period can be derived independently from the DDE system \cite{kiss2015pairwise}, and they coincide with the result of taking the limit as $K \to \infty$ for the multi-stage model (\ref{eq:ex_pair}). This model, therefore, bridges the gap between the traditional Markovian and delay-based scenarios, and accurately represents the spread of a disease with a distribution of infectious period which cannot be modelled by either. 

\section{The pairwise model on clustered networks}
\label{sec:clustered}

As has already been mentioned, clustering is known to play an important role in the spread of epidemics on networks. A convenient way to quantitatively characterise the level of clustering in a given network is through the clustering coefficient $\phi$, most commonly defined as the proportion of closed triangles of nodes out of the total number of triples (open and closed together) in the network. This coefficient can be computed as follows \cite{keeling1999}
\begin{equation}
	\phi = \frac{\text{trace}(A^3)}{||A^2|| - \text{trace}(A^2)},
	\label{eq:cluster}
\end{equation}
where $A=(a_{ij})_{i,j=1,2,\dots,N}$ is the adjacency matrix of the network, with $a_{ij}=a_{ji}$, $a_{ii}=0$ for all $i,j$, $a_{ij}=1$ if nodes $i$ and $j$ are connected and zero otherwise, and $||\cdot||$ stands for the sum of all the elements of the matrix. In the previous section it was assumed that $\phi = 0$. The challenge presented by clustered networks is that one can no longer assume that all triples are open, and, therefore, the closures of pairwise models have to be reconsidered and appropriately modified to effectively approximate the dynamics. In the most general formulation, one can start from a triple $[X_aS_bY_c]$ where the degree of nodes is considered explicitly. Based on \cite{house2011epidemic}, we can write
\begin{equation}
[X_aS_bY_c] \approx \frac{b-1}{b} \frac{[X_aS_b][S_bY_c]}{[S_b]}\left(1-\phi + \phi\frac{n_1N}{ac}\frac{[X_aY_c]}{[X_a][Y_c]}\right), 
\label{eq:standard_het}
\end{equation}
where again $X,Y \in \mathbb{S}$. In order to remove the dependency on node degree, we employ two \emph{a priori} approximations first introduced by~\cite{eames2002modeling}. The first of these approximations has already featured earlier in~(\ref{eq:secondapprox}), namely,
\[
[X_aY] \approx \frac{a[X_a]}{\sum_j j [X_j]}[XY],
\]
and the second has the form
\begin{equation}\label{eq:firstapprox}
[X_aY_b]\approx \frac{[X_aY][XY_b]}{[XY]}\frac{[ab]n_1 N}{a[a]b[b]}\approx \frac{[X_aY][XY_b]}{[XY]},
\end{equation}
where $n_1$ is the mean degree, and $[a]$ is the expected number of individuals with degree $a$ in the network. The new approximation assumes that the joint probability of a pair can be accurately estimated by removing dependence on the degree of the second node and multiplying by a second term that captures the specifics of the network structure. This term is known as the \emph{assortativity} of nodes with degrees $a$ and $b$, and it measures whether nodes with similar degrees are more likely or less likely to connect to each other \cite{newman2003mixing}. The simplification shown in (\ref{eq:firstapprox}) assumes null assortativity (i.e. random connection between nodes) and will be used throughout this section.

We are now in a position to derive closures for the multi-stage model on clustered networks. In (\ref{eq:standard_het}) the terms outside the bracket are similar to the closure in (\ref{eq:new_close}) and (\ref{eq:zeta}) for unclustered networks. In fact, the sum over all degrees $a$ and $c$ will result in the same expression but with the subscripts dropped, as can be checked using (\ref{eq:secondapprox}) and (\ref{eq:firstapprox}). Thus, the first part of the derivation follows exactly the same methodology as for the unclustered network case discussed in Section~\ref{sec:basic}. Focusing on the final term in (\ref{eq:standard_het}), which is responsible for clustering, we use the above approximations to obtain
\begin{align*}
\frac{n_1N}{ac}\frac{[X_aY_c]}{[X_a][Y_c]} &\approx \frac{n_1N}{ac} \frac{[X_aY]}{[XY][X_a]}\frac{[XY_c]}{[Y_c]}  \approx \frac{n_1N}{ac} \frac{a[X_a][XY]}{[XY][X_a]\sum_i i[X_i]}\frac{[XY_c]}{[Y_c]},\\
&\approx \frac{n_1N}{c} \frac{1}{\sum_i i[X_i]}\frac{c[Y_c][XY]}{[Y_c]\sum_j j[Y_j]}\approx {n_1N} \frac{[XY]}{\sum_i i[X_i]\sum_j j[Y_j]}.
\end{align*}
In a similar way as it was done for $T_1$, it is possible to define $J_1^{(i)}$ and $P_1$ as the sums of all edges emanating from infected nodes in the $i$-th stage and from removed nodes, respectively. Then $J_1 = \sum_{j=1}^K J_1^{(j)}$ is the number of edges pointing outwards from \emph{all} infected nodes, regardless of their degree and the stage of the disease which they are in. The full closures necessary for the model with clustering can now be stated as follows,
\begin{equation}\label{eq:new_close_clust}
\begin{array}{l}
\displaystyle{[SSI] = \zeta(t) \frac{[SS][SI]}{[S]}  \left( 1-\phi + \phi n_1N\frac{[SI]}{T_1J_1}\right),}\\\\
\displaystyle{{[ISI^{(i)}]} = \zeta(t) \frac{[IS][SI^{(i)}]}{[S]}  \left( 1-\phi + \phi n_1N\frac{[II^{(i)}]}{J_1J_1^{(i)}}\right), \quad \text{for} \quad i = 1, 2, \ldots, K,}\\\\
\displaystyle{{[ISR]} = \zeta(t)\frac{[SI][SR]}{[S]}  \left( 1-\phi + \phi n_1N\frac{[IR]}{J_1 P_1}\right),}\\\\
\end{array}
\end{equation}
where $\zeta(t)$ is still given by~(\ref{eq:zeta}), and we have defined
\begin{align*}
T_1 &= [SS] + \sum_{i=1}^{K}[SI^{(i)}] + [SR], \\
J_1^{(j)} &= [SI^{(j)}] + \sum_{i=1}^{K}[I^{(i)}I^{(j)}] + [I^{(j)}R], \qquad J_1 = \sum_{j=1}^K J_1^{(j)},\\
P_1 &= [SR] + \sum_{i=1}^{K}[I^{(i)}R] + [RR].
\end{align*}
The model now has to explicitly consider every possible combination of pairs, which, for a disease with a $K$-stage gamma distributed infectious period, yields the following system of $(K^2 +3K+4)$ equations
\begin{align}
&\dot{[S]}   = -\tau [SI], \nonumber\\
&\dot{[I^{(1)}]} = \tau [SI] - K\gamma [I^{(1)}], \nonumber\\
&\dot{[I^{(j)}]} = K\gamma [I^{(j-1)}] - K\gamma [I^{(j)}], \quad \text{for} \quad j= 2, 3, \ldots, K, \nonumber\\
&\dot{[SS]}= -2\tau[SSI], \nonumber\\
&\dot{[SI^{(1)}]}= -(\tau+K\gamma)[SI^{(1)}] + \tau\left([SSI] - [ISI^{(1)}]\right), \nonumber\\
&\dot{[SI^{(j)}]}= -(\tau+K\gamma)[SI^{(j)}] + K\gamma[SI^{(j-1)}] -\tau [ISI^{(j)}], \quad \text{for} \quad j = 2, 3, \ldots, K, \nonumber\\
&\dot{[SR]} = -\tau [ISR] + K\gamma [SI^K], \label{eq:ex_pair_cluster} \\
&\dot{[I^{(1)}I^{(1)}]} = 2\tau [SI^{(1)}] + 2\tau [ISI^{(1)}] - 2K\gamma[I^{(1)}I^{(1)}], \nonumber\\
&\dot{[I^{(1)}I^{(j)}]} = \tau[SI^{(j)}] + \tau [ISI^{(j)}] + K\gamma\left( [I^{(1)}I^{(j-1)}] -2 [I^{(1)}I^{(j)}] \right), \quad \text{for} \quad j= 2, 3, \ldots, K, \nonumber\\
&\dot{[I^{(j)}I^{(k)}]} = K\gamma\left( [I^{(j-1)}I^{(k)}] + [I^{(j)}I^{(k-1)}] -2 [I^{(j)}I^{(k)}] \right), \quad \text{for} \quad j, k= 2, 3, \ldots, K, \nonumber\\
&\dot{[I^{(1)}R]}	= \tau [ISR] + K\gamma\left([I^{(1)}I^{(K)}] - [I^{(1)}R]\right), \nonumber\\
&\dot{[I^{(j)}R]}   = K\gamma\left([I^{(j)}I^{(K)}] + [I^{(j-1)}R] - [I^{(j)}R]\right), \quad \text{for} \quad j= 2, 3, \ldots, K, \nonumber\\
&\dot{[RR]}       = 2K\gamma [I^{(K)}R],\nonumber
\end{align}
with the closures for $[SSI]$, $[ISI^{(j)}]$ and $[ISR]$ given in (\ref{eq:new_close_clust}). Note that, as one would expect, setting $\phi=0$ reduces this model back to the simpler compact model introduced and discussed in Section \ref{sec:basic}.

\begin{figure}
	\hspace{-1cm}
	\epsfig{file=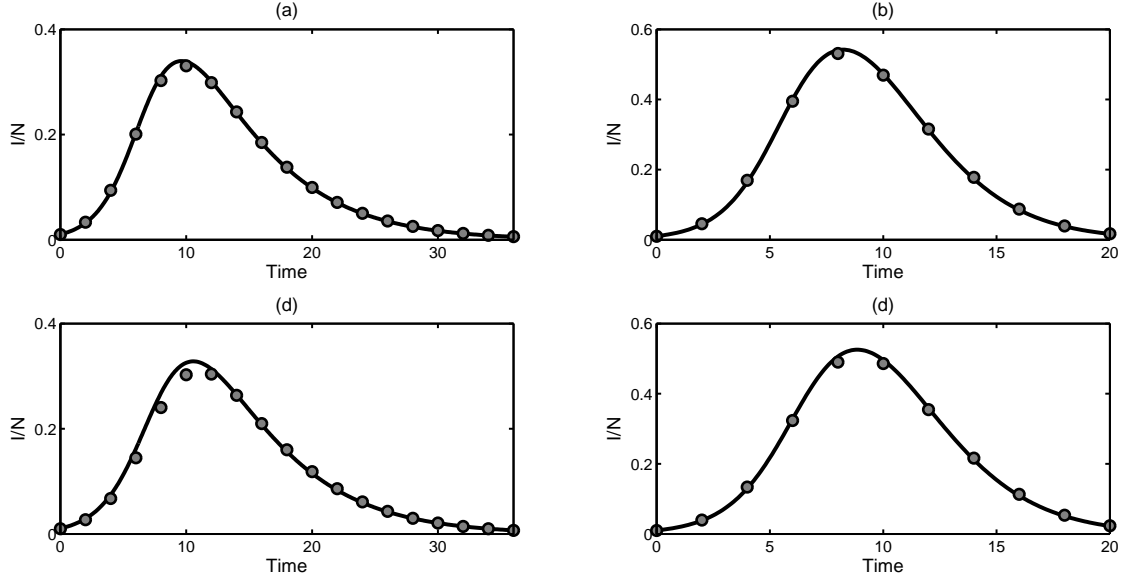,width=18cm}
	\caption{Comparisons of numerical results (circles) and the compact pairwise model on a bimodal network with an even split of nodes having degrees $4$ and $12$. (a) $K=1$, $\phi=0$, (b) $K=5$, $\phi=0$, (c) $K=1$, $\phi=0.2$, (d) $K=5$, $\phi=0.2$. Other parameter values are $\tau$=0.1, $\gamma=0.2$.}
	\label{fig:clust_bimod}
\end{figure}

\subsection{Numerical Simulations}

To investigate the accuracy of model~(\ref{eq:ex_pair_cluster}), we compare its output to stochastic network simulation. First, it is necessary to explain how one can construct clustered networks, which is achieved using the \emph{big-V rewiring} method \cite{bansal2009exploring}. This algorithm takes as an input a random unclustered network constructed with the configuration model, and at each iteration it looks for a chain of five nodes $u - v - x - y - z$, such that newly created links after the rewiring process do not yet exist. Once such a chain is found, the algorithm deletes $u - v$ and $y - z$ edges, and connects $v - y$ and $u-z$ in order to replace the five-node chain with a triangle and a separately connected edge. If this procedure increases local clustering, then the rewiring is accepted, and the algorithm continues until the target clustering coefficient $\phi$ is reached. The benefit of this approach is that while the level of clustering can be varied, the degree distribution remains the same.

Figure \ref{fig:clust_bimod} illustrates the results of simulations on bimodal networks both for unclustered networks and for rewired networks with the clustering coefficient $\phi=0.2$. Whilst the agreement is good in all cases, the clustering introduces some inaccuracy. This is to be expected since the number of susceptible neighbours of a node is now harder to predict due to the presence of short cycles. Furthermore, the inclusion of triangles appears to slow down the spread of the epidemic. The grouping of nodes into small communities decreases the number of individuals at risk of infection at any time, because the disease has fewer routes to spread away from an infectious seed. One should also note that with the introduction of a gamma-distributed infectious period, the trend of faster epidemic growth and higher peak prevalence with increasing values of $K$ is preserved. This reinforces the earlier conclusion that the inclusion of a more realistic distribution of infectious periods can lead to more rapid severe epidemics than what would be predicted by the traditional models with an exponentially distributed infectious period.

\begin{figure}
	\hspace{-1cm}
	\epsfig{file=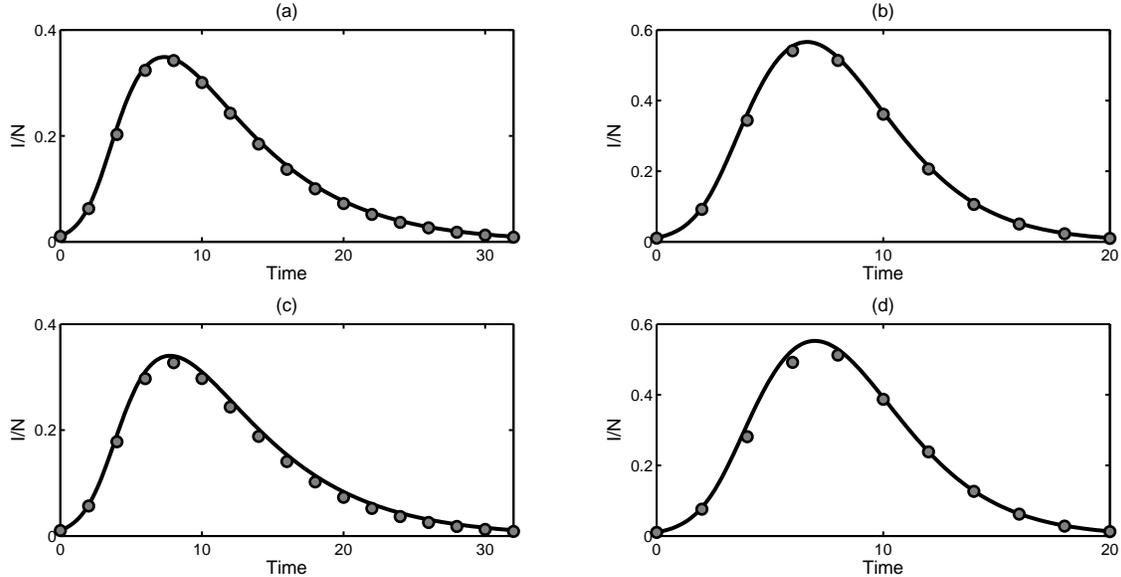, width=18cm}
	\caption{Numerical simulations (circles) compared to the pairwise model (black line) for truncated scale-free networks with exponent $\alpha=2.5$, $\tau$=0.1, $\gamma=0.2$. (a) $K=1$, $\phi=0$; (b) $K=5$, $\phi=0$; (c) $K=1$, $\phi=0.2$; (d) $K=5$, $\phi=0.2$.}
	\label{fig:SF_cluster_change}
\end{figure}

Similar changes in the dynamics are observed in the case of truncated scale-free networks, as shown in Fig.~\ref{fig:SF_cluster_change}. However, unlike the bimodal case, the impact of higher clustering has a less pronounced effect on the timescale of the epidemic. This is likely due to the fact that highly connected nodes cannot be effectively restricted to a single small community, and, therefore, their ability to spread the disease is not significantly affected. It can also be seen that a larger value of $K$ appears to improve the accuracy of the pairwise model~(\ref{eq:ex_pair_cluster}).
 
\begin{figure}[h]
	\epsfig{file=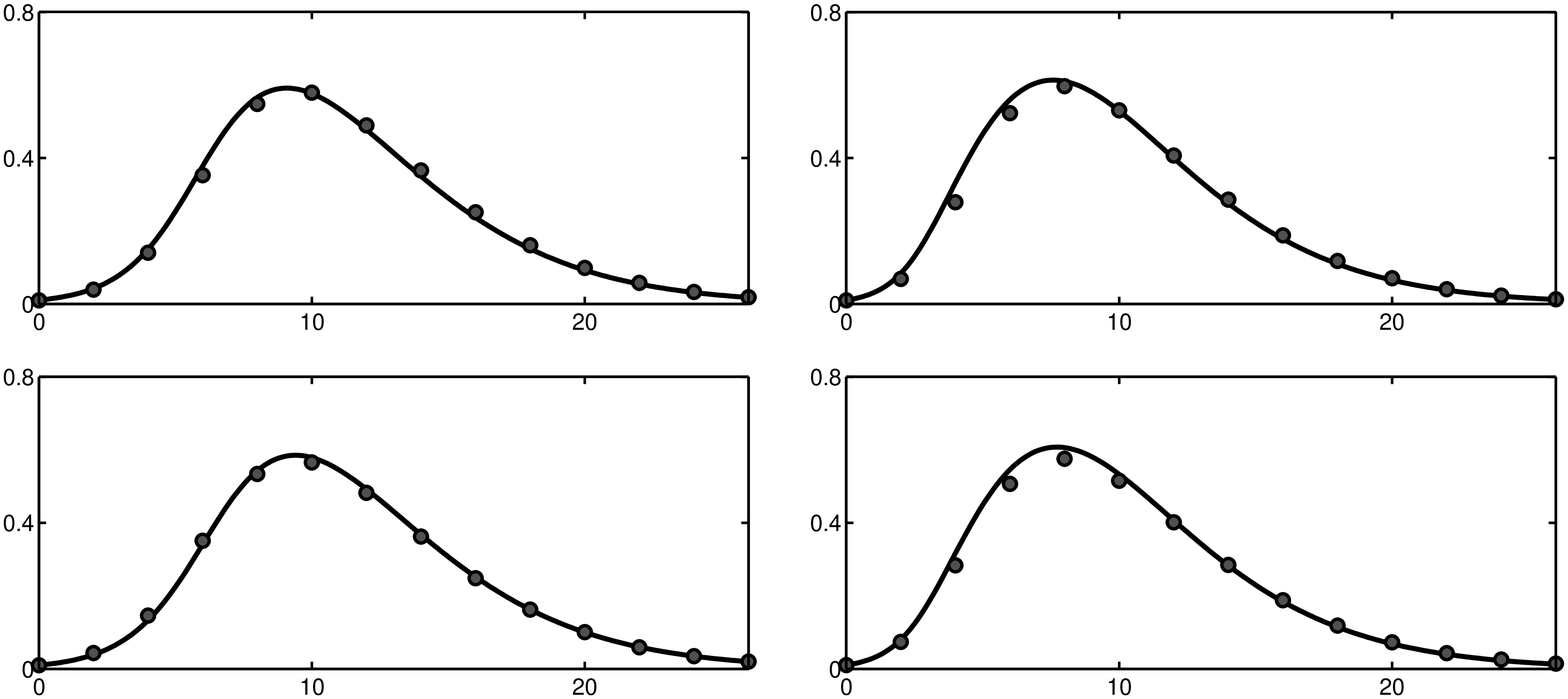, width=15cm}
	\epsfig{file=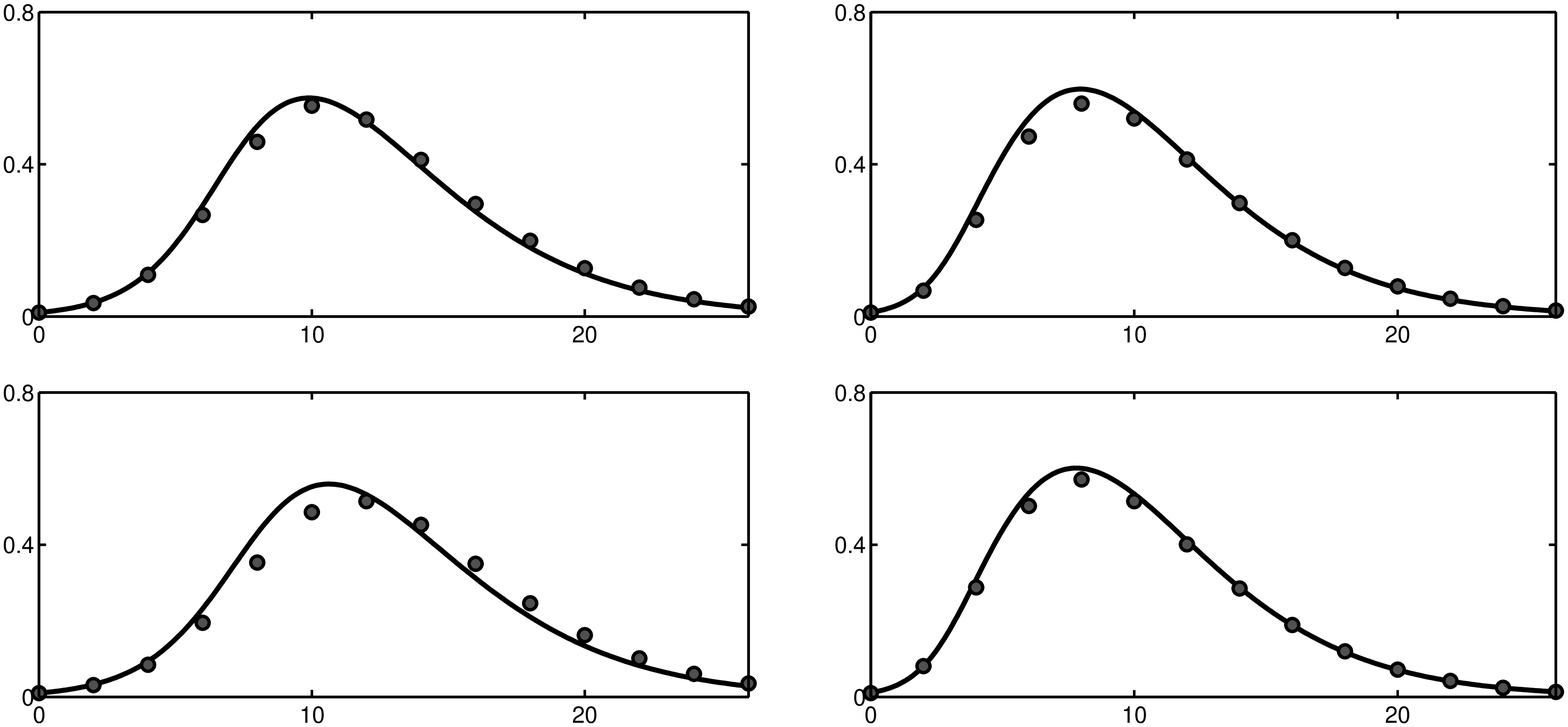, width=15cm}
	\caption{Comparison between the clustered pairwise model (\ref{eq:ex_pair_cluster}) (solid lines) and network simulation (circles) for different values of the clustering coefficient $\phi$. The left column shows results for bimodal networks with an even split of nodes having degrees $4$ or $12$, the right column shows the results for truncated scale-free networks with the exponent $\alpha = 2.5$ and node degrees bounded between $k_{min}=4$ and $k_{max}=60$. Parameter values are $\tau=0.1$, $\gamma=0.15$, $K=3$, with $\phi$  increasing through $0.1$, $0.2$, $0.3$, $0.4$ from top to bottom.  As the clustering $\phi$ increases beyond $0.2$, the inaccuracy of the clustered pairwise model becomes more pronounced.}
	\label{fig:increasing_phi}
\end{figure}
\begin{figure}
	\hspace{-1cm}
	\epsfig{file=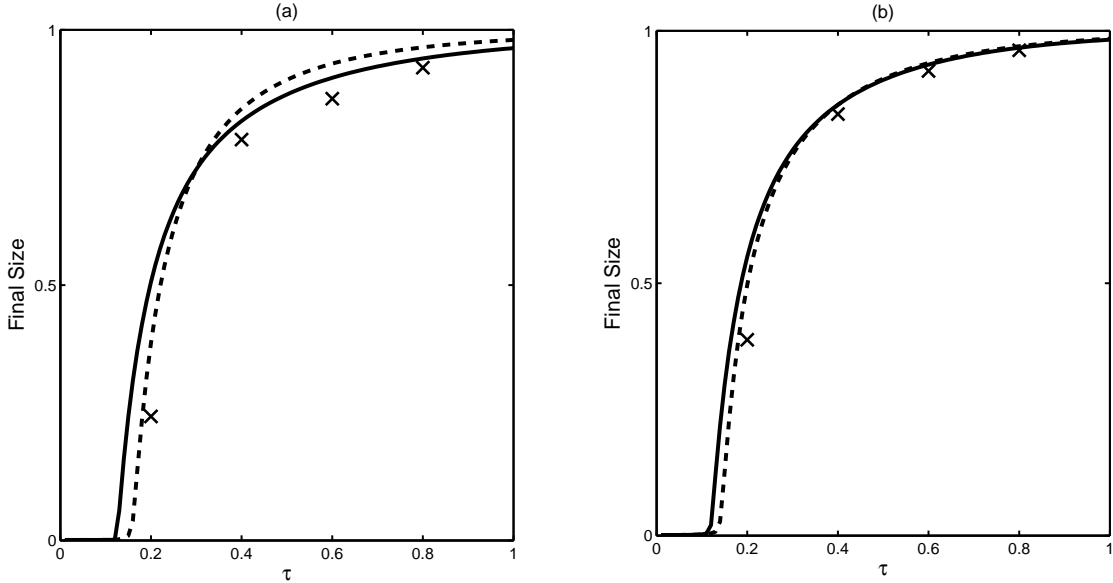, width=18cm}
	\caption{Dependence of the final epidemic size on the per-link transmission rate for bimodal networks with degree 4 or 12, split equally. The parameter $\gamma$ is fixed at 1, and in (a) $K=1$, (b) $K=3$. Solid lines correspond to epidemics on unclustered networks, dashed lines illustrate equivalent epidemics on a network with $\phi=0.4$, and the crosses represent results from 100 numerical simulations on the same clustered networks.}
	\label{fig:inaccuracies}
\end{figure}

Despite its successes, the pairwise model~(\ref{eq:ex_pair_cluster}) becomes less accurate as clustering in the network increases. To investigate this in more detail, we have performed numerous comparisons between simulations and the numerical solution to the pairwise model (\ref{eq:ex_pair_cluster}) for networks with bimodal and truncated scale-free degree distributions, with increasing levels of clustering. The results of these tests are presented in Fig.~\ref{fig:increasing_phi} which shows that system~(\ref{eq:ex_pair_cluster}) is reasonably accurate for low levels of $\phi$, however, this accuracy reduces as $\phi$ increases. The most likely explanation for this reduction in model accuracy is the assumption of null assortativity explicitly made in ~(\ref{eq:firstapprox}) when deriving closures for the clustered model, since it is known that clustering in networks increases assortativity \cite{foster2011clustering}. Furthermore, it has also been shown in a number of earlier studies that high levels of assortativity are the norm in real social networks (see, for example, \cite{newman2002assortative}). Since the null assortativity assumption is violated in such networks, it is not surprising that the pairwise model (\ref{eq:ex_pair_cluster}) does not provide an accurate representation of dynamics for high levels of clustering. Figure~\ref{fig:inaccuracies} shows the comparison in terms of the final epidemic size recorded from simulation and the pairwise model. Again, it is clear to see that the pairwise model performs less well for higher levels of clustering. However, what can be seen from the results is that when clustering is present in the network the threshold appears to increase and thus measurable epidemics are less likely to occur. This can be seen in Fig.~\ref{fig:inaccuracies}, as a higher transmission rate is required in order for the final epidemic size to diverge away from zero when the epidemic takes place in a clustered network. Similarly, for clustered networks, simulation results show that the final epidemic size will be reduced when compared to equivalent networks with the same degree distribution, no clustering and the same parameters of the disease dynamics. This makes sense intuitively, since rewiring a network makes the population more segregated and thus less at risk of widespread epidemics.

Figure \ref{fig:inaccuracies} further suggests that the difference between the pairwise model and simulations is less marked for the non-Markovian case (i.e. $K>1$). In an extensive recent study of small/toy networks, \cite{pellis2015exact} proved that for an $SIR$ epidemic on a single open triple or closed triangle the classical closures, such as those given in (\ref{eq:mid_closure}) and (\ref{eq:standard_het}), are exact for constant infectious periods (see Proposition 3 in \cite{pellis2015exact}). As has been previously discussed in Section~\ref{sec:Kinf}, as the number of stages, $K$, increases in the pairwise model, we approach the limit of a constant infectious period. Therefore, if the results of~\cite{pellis2015exact} extend to larger networks, one would expect that the accuracy of our pairwise
\newpage

\begin{figure}
	\hspace{-1cm}
	\epsfig{file=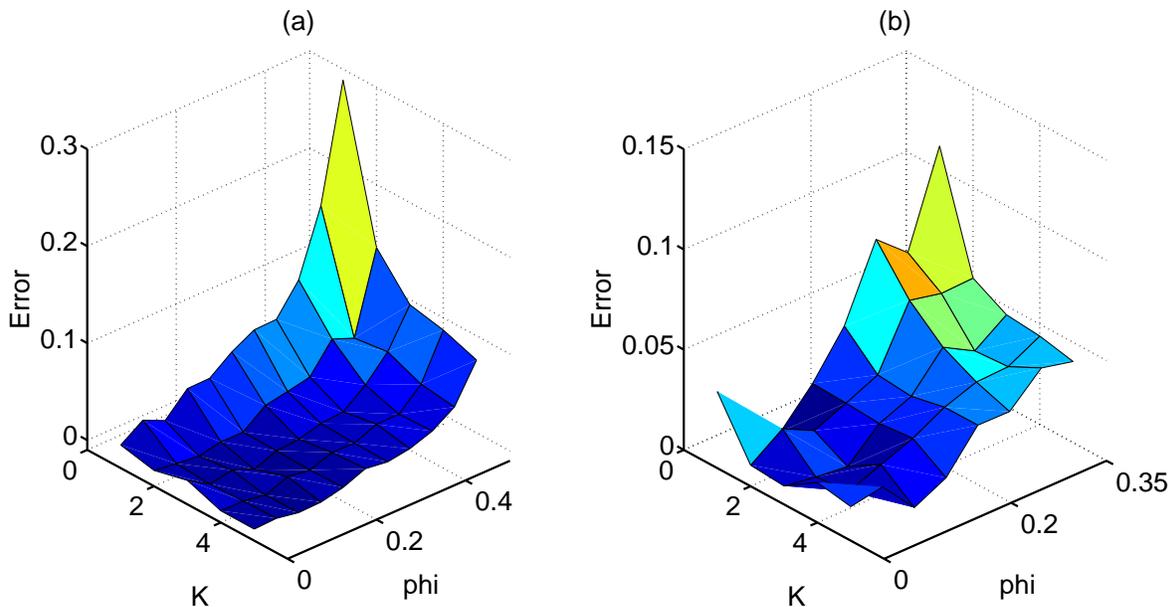, width=18cm}
	\caption{The error between the final epidemic sizes obtained from the solution of the pairwise model (\ref{eq:ex_pair_cluster}) and from the average of 100 numerical simulations plotted against the clustering coefficient $\phi$ and the number of stages of infection $K$. (Plotted as pairwise subtract simulation). Parameter values are $\tau = 0.3$, $\gamma = 1$.  (a) A bimodal network with an even split of nodes having degrees $k_1=4$ and $k_2 = 12$. (b) A scale-free network of 1000 nodes with $k_{min} = 4$, $k_{max} = 60$, and $P(k) \sim k^{-2.5}$. Note that as predicted, even in the presence of clustering, as $K$ grows the error becomes smaller, and hence the pairwise model becomes more accurate.}
	\label{fig:error_surface}
\end{figure}

\noindent model for clustered networks (\ref{eq:ex_pair_cluster}) should improve as $K$ increases. To test the validity of this hypothesis, in Fig.~\ref{fig:error_surface} we plot the value of the error between the final epidemic size computed from the pairwise model~(\ref{eq:ex_pair_cluster}) and the results of 100 simulations under the same parameters, for bimodal and truncated scale-free networks. Figure \ref{fig:error_surface} indicates that the error does indeed decrease for any $\phi$ as the infectious period becomes tighter around the mean (as characterised by an increasing $K$), thus providing evidence that Pellis et al.'s results are relevant for large networks where both open and closed triples are present. Furthermore, one should note that in all but two cases the pairwise model~(\ref{eq:ex_pair_cluster}) over-estimates the final epidemic size when compared to simulations. This suggests that in most cases the model can be expected to give an upper bound on the size of an epidemic. Even though our newly derived compact closures (\ref{eq:new_close_clust}) are not exact, their performance improves greatly when the infectious period approaches the limit of a fixed infectious period. This is an important result that justifies the continued use of pairwise-like methods for non-Markovian epidemics on networks.

\section{Discussion}
\label{sec:postmortem}

In this paper we have derived and studied a new pairwise model for the spread of infectious diseases which includes three major characteristics that are not consistently studied concurrently, despite being essential for understanding disease dynamics in many realistic scenarios. Our pairwise model can account for degree heterogeneity, clustering and gamma-distributed infectious periods, and the number of equations in the pairwise model does not depend on the range of different node degrees. This approach follows the methodology of the so-called \emph{compact pairwise models} \cite{simon2015super, house2011insights}, and the output from the resulting pairwise model shows excellent agreement with results of numerical simulation for networks with either no or low levels of clustering, and for all the different degree distributions that have been considered.

In the absence of clustering we have used linear stability analysis to determine a threshold parameter from the pairwise model, and we have shown that existing methods for finding the final epidemic size \cite{newman2002spread} can be applied. Equivalent results have not been found in the case of clustered networks. However, extensive numerical simulations have shown that introducing multiple stages of infection increases the speed of epidemic spread, as well as the peak prevalence and the final epidemic size. The interactions of degree heterogeneity, clustering and the distribution of infectious period all have significant yet contrasting impacts on an outbreak. For example, we have seen that both degree heterogeneity and a larger number of infectious stages (corresponding to a tighter distribution for the duration of infection) increase the growth rate of the epidemic in the early stages, however, this is countered when one includes clustering that is likely to be present in real contact networks. These findings are consistent with earlier results on the effects of clustering on the spread of epidemics \cite{eames2008modelling}. \cite{serrano2006percolation} have shown that whilst clustering makes epidemics less likely, for scale-free topologies, and in the limit of infinite networks, an epidemic threshold does not exist, and a significant outbreak will always occur. The complexity of the pairwise model for clustered networks has meant that analytical expressions for the epidemic threshold and the final epidemic size have not been found. In fact, analytical results have so far only been obtained for clustered networks with a specific construction, e.g. non-overlapping triangles \cite{miller2009percolation}. Random rewiring enforces fewer restrictions on the network and thus allows for more complex topologies to emerge; it is likely to provide more realistic but also more challenging scenarios for modelling than networks with a prescribed nature of clusters.

A strength of the final pairwise model which we have presented is that it can be tuned based on the characteristics of the disease and population being studied. There are several ways to include more features into this model. For example, assortativity could be made an explicit consideration in the closures, and by allowing the transmission rate to vary depending on the stage of infection, one could model diseases with varying infectivity. Setting $\tau=0$ in any number of initial stages also opens the possibility for multi-stage $SEIR$ models to be studied, again without altering the basic framework of the model.

Models, such as the on presented in this paper, could also be used for a more thorough study of the performance of closures and for mapping out how different approximations behave under different regimes, such as stochastic models for the transmission and recovery processes. Furthermore, one could consider whether non-Markovian transmission processes can be incorporated into pairwise or pairwise-like models. Additional motivation for research into this area comes from studies which have suggested that human contact patterns are typically very 'bursty' \cite{chaintreau2007impact, cattuto2010dynamics}. This means that there are many short periods with high levels of interaction and longer periods of little or no action, and this may have a significant impact on how an epidemic may spread. It is possible that attempts to incorporate non-Markovian transmission may lead to a more complex system of integro-differential equations.

There have been many recent developments in the area of dynamic or adaptive networks \cite{gross2006epidemic, risau2009contact, kiss2012modelling, selley2015dynamic} where pairwise models have been used successfully to couple the dynamics of an epidemic on the network with the dynamics of the network. These models have shown that using pairwise approximation techniques it is possible to capture non-trivial properties of both network and epidemic dynamics in a single model. There is a wide scope for further research focussed on modelling the rewiring process, as well as for analysis of a reaction of networks to a spreading epidemic when considered as a non-Markovian process.

The pairwise model presented in this paper does well at accounting for non-Markovian infectious periods, indeed, it becomes more accurate in this case, yet it is limited in capturing epidemics on realistic clustered networks. This highlights that complexities in the structure of social networks are difficult to model even with a large number of equations. The above suggests that when studying epidemics on networks and designing disease control strategies or interventions, it is essential to use accurate and reliable data about the population being studied, as well as about epidemiological characteristics of the disease.

\section*{Acknowledgements}
Neil Sherborne acknowledges funding for his PhD studies from EPSRC (Engineering and Physical Sciences Research Council), EP/M506667/1, and the University of Sussex.

\section*{References}
\bibliographystyle{ieeetr}
\bibliography{heterogeneous_NS2}

\end{document}